# Electro-kinetics of Polar Liquids in Contact with Non-Polar Surfaces

Chih-Hsiu Lin[1], Gregory S. Ferguson[2] and Manoj K. Chaudhury* [1]

[1] Department of Chemical Engineering and [2] Department of Chemistry,
Lehigh University, Bethlehem, PA 18015

**ABSTRACT:** Zeta potentials of several polar protic (water, ethylene glycol, formamide) as well as polar aprotic (dimethyl sulfoxide) liquids were measured in contact with three non-polar surfaces using closed-cell electro-osmosis. The test surfaces were chemisorbed monolayers of alkyl siloxanes, fluoroalkyl siloxanes and polydimethylsiloxanes (PDMS) grafted on glass slides. All these liquids exhibited substantial electro-kinetics in contact with the non-polar surfaces with these observations: the electrokinetic effect on the fluorocarbon-coated surface is the strongest; and on a PDMS grafted surface, the effect is the weakest. Even though these hygroscopic liquids contain small amounts of water, the current models of charging based on the adsorption of hydroxide ions at the interface or the dissociation of pre-existing functionalities (e.g., silanol groups) appear to be insufficient to account for the various facets of the experimental observations. The results illustrate how ubiquitous the phenomenon of electro-kinetics is with polar liquids contacting such apparently passive non-polar surfaces. We hope that these results will inspire further experimental and theoretical studies in this important area of research that has potential practical implications.

* e-mail: mkc4@lehigh.edu

## 1. INTRODUCTION

It is inferred from electrokinetic measurements that the interfaces of various hydrophobic non-polar solids in contact with pure water become negatively charged. This result is surprising considering the absence of dissociable groups on such solids that could give rise to charging. While it has been stipulated in the literature[1-5] that the adsorption of hydroxide ions is responsible for negative charge at the interface of water and non-polar materials, spectroscopic as well as the theoretical studies[6-12] have largely failed to provide strong evidence for such a model. Although certain MD simulations[13] argued for only a slight accumulation of hydroxide ions at the interface, the preponderance of the evidence from various simulations and spectroscopic studies[9-12], on the other hand, suggest that hydronium ions – not hydroxide ions – accumulate in excess at the interface. Recently, a new effort has been made to explain the charging at the interface of water in contact with hydrophobic surfaces based on donor-acceptor interactions.[14] In this model, partial charge transfer inherent in hydrogen bonds between water molecules gives rise to a zeta potential at the interface, where the isotropic symmetry of the bulk is broken and charge separation is imbalanced. Such an explanation of charging at the water/non-polar interface implies that other types of H-bonding liquids, e.g., ethylene glycol and formamide, should exhibit non-negligible electrokinetic effects as well.

Nearly two decades ago, Yaminsky and Johnston[15] reported that a hydrophobized glass slide acquires negative charge when it is withdrawn from water. In fact, they found that this effect is not restricted to water alone; charging of variable magnitudes is also observed with formamide, mercury and certain solutions of water and ethanol, but not with such pure dielectric liquids as alkanes. The authors speculated that the surfaces acquire charge when a non-wetting liquid breaks its adhesive bonds during the retraction of the solids in a manner akin to triboelectricity.[16] However, for such a charge separation to occur, there has to be a difference in the chemical potential of the electrons on the two surfaces with the possibility of an energy barrier preventing instantaneous charge neutralization during the interfacial separation process. These results then lead to the question: is it possible that the negative zeta potential of water in contact with hydrophobic surfaces is related to a contact-charging phenomenon? The Yaminsky-Johnston experiments should be clearly demarcated from the standard electrokinetic measurements in that the former involves retraction of a solid from a pre-wetted liquid, whereas the latter is performed when the solid remains wetted by a liquid. In order to bridge these two types of experiment, electrokinetic measurements are needed with various polar liquids against non-polar surfaces while they maintain interfacial contact. In this paper, we pursue such a study by measuring the zeta potentials of hydrophobically modified glass slides in contact with polar liquids – ethylene glycol, formamide, and dimethyl sulfoxide – and compare the results with the same surfaces in contact with water. Amongst these liquids, both ethylene glycol and formamide can autoionize only sparingly[17], whereas dimethyl sulfoxide is a non-ionizable aprotic liquid. All of these liquids are however hygroscopic and contain trace amounts of water. Detailed analysis of the electrokinetic data of these liquids, even in the presence of small amounts of water, and how they depend on the chemical nature of hydrophobization suggests that an explanation beyond that of the adsorption of hydroxide ions may be warranted in order to explain the observed charging phenomena.

## 2. EXPERIMENTAL SECTION

### 2.1 MATERIALS

The liquids used for the experiments were deionized water (DI water; Barnstead), ethylene glycol (EG, spectrophotometric grade, 99+%; Alfa Aesar), formamide (FA, spectrophotometric grade, 99+%; Alfa Aesar), and dimethyl sulfoxide (DMSO, HPLC grade, 99.9+%, packaged under argon in re-sealable ChemSeal™ bottles; Alfa Aesar). In one set of experiments, the as-received EG and FA were heated to 120 °C for 2 h under a bubble purge of nitrogen to dry the solvents as much as possible. DMSO was used either straight from the bottle or after vacuum distillation. Both EG and DMSO were distilled under partial vacuum (at ~10 Torr), the latter over calcium hydride (≥97.0%; Sigma-Aldrich). Formamide was used either as received or after equilibrating 100 mL of it with 10 g of mixed-bed resin (cat. No. M8032; Sigma-Aldrich) under a bubble purge of nitrogen for 1 h in order to remove the excess ions. This treatment, however, introduces some amount of water to the solvent. In some experiments, we also added deliberately small amounts of water to the test liquids. Two different fluorescent particles were used in the particle image velocimetry (PIV): FluoSpheres carboxylate-modified microspheres (0.5 μm, yellow-green fluorescent; Invitrogen Co.), and FMY yellow UV fluorescent microspheres (1um-5um; Cospheric), from which the smallest particles were extracted. *n*-Hexadecyltrichlorosilane (HC-16, 95%; Gelest Inc.), *1H,1H,2H,2H*-perfluorodecyltrichlorosilane (FC-10, 96%; Alfa Aesar) and trimethylsiloxy-terminated polydimethylsiloxane (DMS-T22, M.W. 9430; Gelest Inc.) were used as received. The fluid connector (cat. No. 72-1437) and Tygon® laboratory tubing (R-3603, I.D. 1/16 in., O.D. 1/8 in.) were purchased from Harvard Apparatus, Inc.

The amount of water in EG was determined by attenuated total reflectance Fourier transform infrared spectroscopy (ATR-FTIR, Nicolet iS10; Thermo Scientific), using its peak in the 1600-1720 cm$^{-1}$ region and calibrated standards prepared by adding known amounts of water in EG.  The amounts of water in FA and DMSO were determined using nuclear magnetic resonance spectroscopy (NMR, DRX 500; Bruker BioSpin) by dissolving them in DMSO-d6 (D, 99.9%; Cambridge Isotope Laboratories, Inc.) for FA and chloroform-d (D, 99.8%; Cambridge Isotope Laboratories, Inc.) for DMSO, respectively.  Once the in-house spectroscopic measurements were calibrated against the coulometric Karl Fischer titrations (Intertek Pharmaceutical Services, Whitehouse, NJ), the internal measurements were performed for routine analysis of water in the solvents. The pH and the conductivity of the test liquids were measured using a pH meter (model 215 with Gel-filled pH electrode #300737.1; Denver Instrument) and a conductivity meter (model 23226-523; VWR International, LLC), respectively.

For particle velocimetry, all the liquids were seeded with dilute fluorescent particles.  For EG and FA, the FluoSpheres particles were freeze-dried followed by re-dispersion in the test liquids via sonication.  For DMSO, the FMY particles were first washed through a syringe filter (Acrodisc® 25 mm premium syringe filter with 0.45 μm GHP membrane; Pall Life Sciences) with DMSO.  The particles were then flushed back to the syringe (5 mL disposable syringe, model 26214; Exel International) with additional DMSO, and then filtered through Whatman GF/C filter paper a few times to remove the large (>1.2 μm) particles.  The filtrate with concentrated particles was added to the test liquid of DMSO before the electrokinetic measurements.  The surface tensions of all the liquids were measured by the du Noüy ring method (Fisher model 215 autotensiomat surface tension analyzer; Fisher Scientific). To check if any contaminants leached out of the tracer particles, the surface tensions of the liquids were measured before and after they were seeded with the particles. Since the surface tension values did not change, we were assured that there was no detectable contamination.

## 2.2. PREPARATION OF TEST CHANNELS

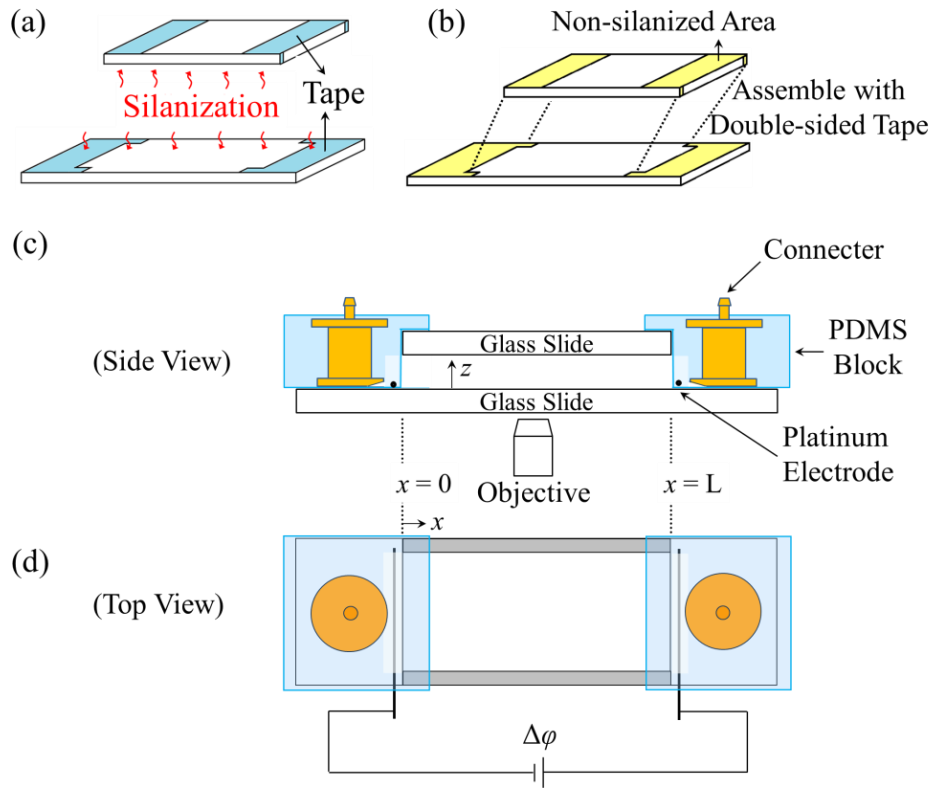

**Figure 1**. Schematic diagrams summarizing the preparation of the testing channel. (a) The two end regions of the glass slides were covered by transparent tape (blue area) to protect the glass surface from silanization. (b) After silanization, the tapes were removed, and the channel was assembled with double-sided tape (gray rectangles in (d)). The yellow areas represent glass surfaces. (c) Side view of the channel. The two ends of the channel were closed with PDMS blocks imbedded with fluid connecters. The platinum electrode was pushed through the PDMS block and placed near the end of the channel. The testing cell was filled with test liquid and placed on the platform of an epifluorescence microscope. (d) Top view of the channel. A voltage ($\Delta\varphi$) was applied across the channel via a power supply. L is the length of the channel.

The assembly of the test channel is shown in Fig. 1. The glass slide ($75 \times 25 \times 1$ mm, cat. No. 12-550-A3; Fisher Scientific) that formed the upper part of the channel was cut to dimensions of $50 \times 25 \times 1$ mm, whereas that forming the bottom of the channel had the as-received dimensions. After both the glass slides were cleaned with piranha solution and dried with a nitrogen purge, the ends of both slides were covered with transparent Scotch® tape as shown in Fig. 1a. The slides were treated with an oxygen plasma (model PDC-32G; Harrick Plasma) and then some of those were allowed to react with the vapor of either HC-16 or FC-10 as reported in the literature.[18,19] The tape that was utilized to protect the end of the glass slides from silanization was removed following the completion of the silanization. Although accurate measurements of thicknesses of the grafted silane layers could not be carried out on glass because the substrate was not extremely smooth, and because the refractive index of the substrate closely matched that of the silane, the thicknesses of equivalent silanized layers on silicon wafers could be characterized using variable angle spectroscopic ellipsometry (model V-VASE® with WVASE32 software; J.A. Woollam Co., Inc.). These measurements gave to a thickness of 2.5 nm for HC-16 and 1.6 nm for FC-10 on an equivalent polished silicon wafer.[20,21]

Some of the glass slides were modified with a thin layer (5 nm) of PDMS (DMS-T22) using a method reported recently in the literature[22]. After the slide was cleaned by piranha solution and then oxygen plasma, it was fully wetted by DMS-T22 and covered with another glass slide placed above it. The sample was kept in a vacuum oven at 80°C for 24 h, followed by cooling to room temperature. The sample was then gently rinsed with distilled chloroform (ACS grade; EMD) and dried with a stream of nitrogen.

Although we have not measured the thickness of the grafted layer of PDMS on glass, similar treatment on a silicon wafer produced the ellipsometric thickness of the adsorbed film of 5 nm.

All of the surfaces were characterized by the contact angles of the test liquids (DI water, ethylene glycol, formamide and dimethyl sulfoxide), atomic force microscopy (AFM, Nanodimension V; Vecco), and x-ray photoelectron spectroscopy (XPS). The contact angles were measured by using the drop inflation and deflation methods. Angles measured on both sides of each drop were averaged from three measurements performed at different spots of each sample. High-resolution XPS scans in the $C_{1s}$ and $Si_{2p}$ regions were carried out at a 15° take-off angle with a SCIENTA ESCA-300 instrument using monochromatic Al $K_\alpha$ x-rays generated from a rotating anode operated at 4.5 kW and a pass energy of 150 eV.

The test channel was prepared by assembling two similar glass slides with double-sided Scotch® tape as shown in Fig. 1b. A PDMS block embedded with a fluid connecter was prepared using Sylgard® 184 (Dow Corning) following the methodology described in the literature.[23] Before assembling the channel, the bottom sides of the PDMS blocks were treated with oxygen plasma at 0.2 Torr for 45 s to enhance adhesion. Clamps were used to affix the plasma-treated PDMS blocks onto the two ends of the channel for 24 h to ensure intimate contact between the PDMS and the surface (Fig. 1c and 1d).

One obvious concern with the prepared channel was the possibility that impurities might leach from the PDMS and the tape used in the assembly process and contaminate the test liquids. In order to ascertain that this was not the case, test liquids were passed through one of the ports of the assembled channel via a Tygon® tube, and then collected through the other port. Once enough liquid was collected, its surface tension was measured and compared with pure liquids. No measurable differences in surface tension ensured that the contamination of the test liquids by putative impurities was not an issue of concern.

## 2.3. MEASUREMENT OF VELOCITY PROFILES OF LIQUIDS

The test liquid seeded with fluorescent particles was injected into the channel through a Tygon® tube and the fluid connecter utilizing a 5-mL disposable syringe. After the test cell was fully filled with the liquid with no visible air bubbles, its fluid connecters were sealed with solid PDMS plugs. As shown in Fig. 1d, platinum electrodes (Premion®, 0.25 mm dia.; Alfa Aesar) were pushed through the PDMS blocks and placed near both the ends of the channel. The test-ready channel was then placed horizontally on the platform of a microscope (model Diaphot; Nikon). When a specific voltage was applied via a power supply (cat. no. 3-1008 from Buchler Instruments or model GPS-1850D from Good Will Instrument Co., Ltd) across the channel through the platinum electrodes, the fluorescent particles began to move, which were observed with the epifluorescence setup (model DM510; Nikon) of the microscope by focusing at different depths of the channel. For each test, electric fields at five different strengths were applied, and the images of particles were recorded on the computer via a CCD camera (model XC-75; Sony).[18]

## 3. THEORY: DETERMINATION OF ZETA POTENTIAL

We begin with the standard equation of electrokinetics[18,24] that balances the viscous, electrical, and normal stresses:

$$\eta \frac{d^2 v(z)}{dz^2} = \varepsilon \varepsilon_0 E \frac{d^2 \psi(z)}{dz^2} + \frac{dp}{dx} \ , \tag{1}$$

where $v(z)$ is the velocity of the fluid moving along the length (x direction) of the channel, $\psi(z)$ is the electrical potential that is uniform in the x direction but varies along the depth of the channel, $\eta$ and $\varepsilon$ are the viscosity and dielectric constant of the test liquid, $\varepsilon_0$ is the dielectric constant in vacuum, and $dp/dx$ and $E$ are the pressure gradient and the electric field, which are uniform along the depth (z direction) but vary across the length of the channel, respectively. These equations are traditionally solved with the following boundary conditions:

$$z = 0 \text{ or } 2h, \ v = v_s, \ \psi = \zeta_w$$

$$z = h, \ dv/dz = 0, \ d\psi/dz = 0 \tag{2}$$

where 2h is the depth of the channel, and $\zeta_w$ is the zeta potential at each of the channel walls that is in contact with the liquid. The liquid can potentially slip against the wall with a slip velocity $v_s$. The general solution of eq. 1 with the help of the boundary conditions given in eq. 2 is as follows:

$$v(H) = \frac{h^2}{2\eta} \frac{dp}{dx} \left(H^2 - 2H\right) - \frac{\varepsilon \varepsilon_0 E}{\eta} \left(\zeta_w - \psi(H)\right) + v_s, \ H = \frac{z}{h} \ . \tag{3}$$

As fluorescent particles are utilized as tracers in the system and each particle has its own electrophoretic velocity ($v_p$), the velocity profile of the liquid as measured ($v_{exp}$) with the tracer particles is the superposition of v and $v_p$, i.e., $v_{exp} = v + v_p$. The electrophoretic velocity of the particle is same as the depth average value of v, i.e.,

$$v_{exp}(H) = \frac{h^2}{2\eta} \frac{dp}{dx} \left(H^2 - 2H\right) - \frac{\varepsilon \varepsilon_0 E}{\eta} \left(\zeta_w - \psi(H)\right) + v_s + \frac{1}{2} \int_0^2 v_{exp}(H) dH. \tag{4}$$

Following the standard protocol, the slip velocity ($v_s$) of the test liquids against the walls of the flow channel can be expressed[25] as

$$v_s = b \frac{dv}{dz}\Big|_{z=0} = b\left(\frac{\varepsilon\varepsilon_0 E}{\eta}\frac{d\psi}{dz}\Big|_{z=0} - \frac{h}{\eta}\frac{dp}{dx}\right). \tag{5}$$

Here, $b$ is the slip length, and $\frac{dv}{dz}\Big|_{z=0}$ is the velocity gradient at the wall. By integrating eq. 1, the electrical potential gradient $\left(\frac{d\psi}{dz}\Big|_{z=0}\right)$ at the wall can be obtained (eq. 8) from the solution of the Poisson-Boltzmann equation (eq. 6, for a symmetric electrolyte):

$$\frac{d^2\psi(z)}{dz^2} = \frac{2qen_\infty}{\varepsilon\varepsilon_0}\sinh\left(\frac{qe\psi(z)}{k_BT}\right) = \frac{\kappa^2}{A}\sinh\left(A\psi(z)\right), \tag{6}$$

with

$$A = \frac{qe}{k_BT} \text{ and } \kappa = \left(\frac{2n_\infty q^2 e^2}{\varepsilon\varepsilon_0 k_BT}\right)^{1/2}, \tag{7}$$

where $q$ is the charge of the ion, $e$ is the elementary charge, $n_\infty$ is the number of the ions per unit volume, $k_B$ is the Boltzmann constant, $T$ is room temperature (296 K), and $\kappa$ is the Debye-Hückel parameter.

$$\frac{d\psi}{dz}\Big|_{z=0} = \frac{\kappa}{A}\sqrt{2[\cosh(A\zeta_w)-1]}. \tag{8}$$

Now, combining eqs. 4, 5 and 8, we have the velocity profile in the channel as follows:

$$v_{exp}(H) = \frac{dp}{dx}\left[\frac{-2bh + h^2(H^2 - 2H)}{2\eta}\right] + \frac{\varepsilon\varepsilon_0 E}{\eta}\frac{b\kappa}{A}\sqrt{[\cosh(A\zeta_w)-1]} - \frac{\varepsilon\varepsilon_0 E}{\eta}(\zeta_w - \psi(H)) + \frac{1}{2}\int_0^2 v_{exp}(H)dH, \tag{9}$$

where, the surface potential[26,27] has the following form:

$$\psi(\text{H}) = \frac{4}{A}\left\{\tanh^{-1}\left[\tanh\left(\frac{A}{4}\zeta_w\right)\exp(-\kappa h\text{H})\right] + \tanh^{-1}\left[\tanh\left(\frac{A}{4}\zeta_w\right)\exp(-\kappa h(2-\text{H}))\right]\right\} \quad (10)$$

The velocity profile of a particle at a depth H can be obtained experimentally and fitted with eq. 9 to obtain the zeta potential ($\zeta_w$), provided that the slip length (*b*) and the pressure gradient (*dp/dx*) are known.

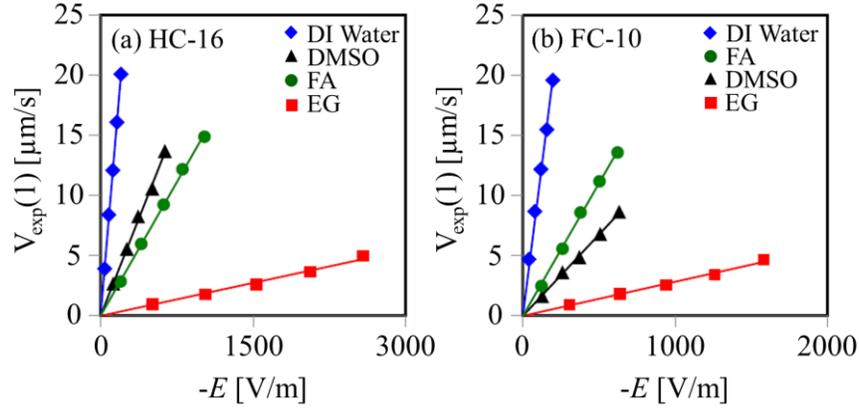

**Figure 2.** Centerline (H=1) velocity of the tracer particles in the channel as a function of electric field strength (-*E*). These (a) HC-16 and (b) FC-10 silane-treated surfaces were tested with four different liquids: DI water, dimethyl sulfoxide (DMSO), formamide (FA), and ethylene glycol (EG).

Various studies in the literature suggest that the value of *b* is in the range of 20 nm or less.[25,28] Thus, one approach would be to take reasonable values of *b* from the literature and fit the velocity profiles to determine under what conditions a best fit is obtained. In order to increase the reliability of such a fit, we measured velocity profiles of the tracer particles at five different values of the applied field *E* (Fig. 2). In all cases, we found that the centerline velocity of the tracer particles is linearly proportional to the electric field, suggesting that the pressure gradient *dp/dx* has to be proportional to *E*. This observation is in agreement with our previous publication[18], in which we devised a way to measure the centerline velocity of the tracer particles in both the presence and the absence of an electric field. In the latter case, the flow still occurred, as the electro-osmotically generated pressure gradient was stored via capillarity. It is thus reasonable to set the pressure gradient in the channel to be a linear function of the electric field:

$$\frac{dp}{dx} = D\varepsilon\varepsilon_0 E, \qquad (11)$$

where $D$ is a constant. Now, dividing the velocity $v_{exp}$ by the electric field and dielectric constant, and multiplying it by the viscosity, we obtain an expression for the mobility[29] ($\tilde{V}_{exp}(H)$) of the fluid as:

$$\tilde{V}_{exp}(H) = -\frac{\eta v_{exp}}{\varepsilon\varepsilon_0 E} = \frac{D}{2}\left[2bh - h^2\left(H^2 - 2H\right)\right] - \frac{b\kappa}{A}\sqrt{2\left[\cosh\left(A\zeta_w\right) - 1\right]} + \left(\zeta_w - \psi(H)\right) + \left\langle\tilde{V}_{exp}\right\rangle, \qquad (12)$$

where $\left\langle\tilde{V}_{exp}\right\rangle$ is the depth average value of $\tilde{V}_{exp}$, which is related to the electrophoretic mobility ($\mu_e$) of the tracer particle:

$$\mu_e = -\frac{v_p}{E} = \frac{\varepsilon\varepsilon_0}{\eta}\left\langle\tilde{V}_{exp}\right\rangle. \qquad (13)$$

## 4. RESULTS AND DISCUSSION

### 4.1. CHARACTERIZATION OF SILANIZED SURFACES

The test surfaces were characterized by atomic force microscopy, wettability, and the x-ray photoelectron spectroscopy (XPS). The corresponding results are summarized in Table 1, Table 2 and Figure 3, respectively. The values of the root mean square (RMS) roughness collected by AFM over an area of 2 μm × 2 μm indicate that the surfaces of the glass slides are relatively smooth, but that they are rough enough to (potentially) prevent slippage of the liquids at the wall.[28]

**Table 1**. The root mean square (RMS) roughness of the silanized glass surfaces over an area of 2 μm × 2 μm.

| | Glass slide (pre-treated with piranha solution) | HC-16 modified | FC-10 modified |
|---|---|---|---|
| RMS roughness (nm) | $0.60 \pm 0.15$ | $0.60 \pm 0.10$ | $1.04 \pm 0.35$ |

**Table 2**. The contact angles of the test surfaces with different liquids. $\theta_a$: advancing contact angle; $\theta_r$: receding contact angle; $\theta_h = \theta_a - \theta_r$.

| | | HC-16 modified | | | FC-10 modified | | | grafted PDMS | | |
|---|---|---|---|---|---|---|---|---|---|---|
| Liquid | Molar Volume (cm$^3$/mol) | $\theta_a$ (°) | $\theta_r$ (°) | $\theta_h$ (°) | $\theta_a$ (°) | $\theta_r$ (°) | $\theta_h$ (°) | $\theta_a$ (°) | $\theta_r$ (°) | $\theta_h$ (°) |
| DI Water | 18.0 | $110 \pm 1$ | $96 \pm 1$ | 14 | $117 \pm 2$ | $93 \pm 1$ | 24 | $103 \pm 2$ | $98 \pm 1$ | 5 |
| EG | 55.7 | $84 \pm 1$ | $76 \pm 1$ | 8 | $99 \pm 1$ | $69 \pm 1$ | 30 | $87 \pm 1$ | $83 \pm 1$ | 4 |
| DMSO | 70.9 | $71 \pm 1$ | $61 \pm 1$ | 10 | $84 \pm 1$ | $70 \pm 1$ | 14 | $78 \pm 1$ | $74 \pm 1$ | 4 |
| FA | 39.7 | $93 \pm 1$ | $83 \pm 1$ | 10 | $107 \pm 1$ | $82 \pm 1$ | 25 | $95 \pm 1$ | $91 \pm 1$ | 4 |

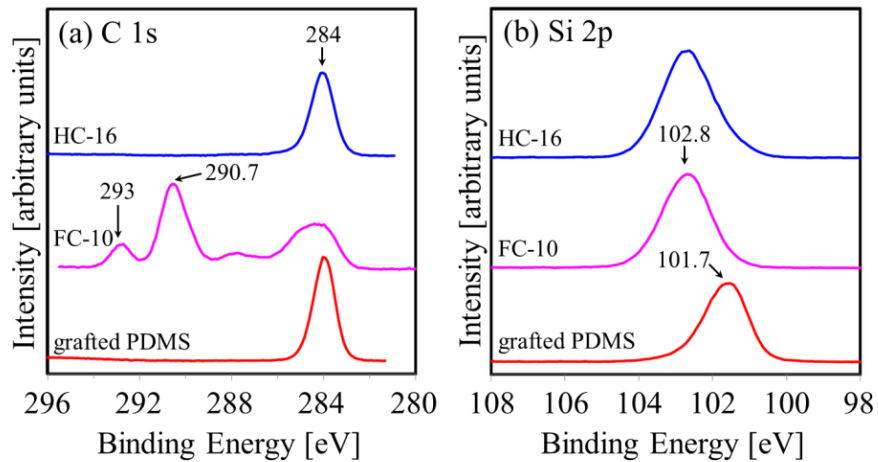

**Figure 3.** X-ray photoelectron spectra (XPS) in the (a) C 1s and (b) Si 2p regions of glass surfaces modified with hydrocarbon silane (blue), fluorocarbon silane (pink) and grafted PDMS (red).

The contact angles of the liquids on these surfaces are comparable to the previously reported values,[18,20,21,30] with the observation that the hystereses of the liquids on the fluorocarbon surface are significantly higher than those on the hydrocarbon surface. These monolayers are very stable in water at neutral pH, as evidenced by the fact that the surface properties of the treated slides remain unaffected even after immersion in pure water for a week. There is a view in the literature[31] that the hysteresis of a test liquid on a surface would systematically decrease with its molar volume. In the current study, however, such a correlation is not observed. For example, ethylene glycol, with a molar volume three times as large as that of water, in fact, exhibits only a slightly higher hysteresis than water, whereas DMSO with an even a larger molar volume than water has a lower hysteresis.

Inspection of the $C_{1s}$ region of the XPS spectra (Fig. 3) reveals that photoemissions from the carbon atoms of the HC-16 and the PDMS surfaces appear only at low binding energy, whereas that from the fluorocarbon surface gives rise to peaks at significantly higher binding energies reflecting the withdrawal of electrons from carbon atoms by the fluorine atoms. Weak photoemission at about 287.4 eV and slight broadening of the peak around 284 eV may indicate incomplete fluorination of the as-received silane. The prominent functionalities are, however, –CF$_3$ (~ 293 eV) and –CF$_2$ (~ 291 eV). Contact angle of water on this surface is insensitive to pH (2 to 12), which is consistent with the absence of ionizable groups such as carboxylic acid. Even though these XPS spectra were collected at a 15° take-off angle, the thin alkyl or fluoroalkyl siloxane monolayers could not fully attenuate the Si$_{2p}$ photoelectrons (103 eV) ejected from the SiO$_2$ of the supporting glass.

## 4.2. ZETA POTENTIALS OF SILANIZED GLASS SLIDES IN VARIOUS LIQUIDS

Fig. 4 summarizes the depth-dependent mobilities [ $\tilde{v}_{exp}(H) - \langle \tilde{v}_{exp} \rangle$ ] of four representative liquids (DI water, ethylene glycol, DMSO and formamide) in contact with the fluorocarbon and hydrocarbon silane-modified glass slides, which are corrected for the mobility of the tracer particles inside the channel (the uncorrected velocity profiles can be found in the Supplementary Materials section). These measurements were carried out at five different electric-field strengths (Fig. 2) for each liquid on each surface. It is gratifying that all five sets of data nicely cluster around a single parabolic velocity profile in each case.

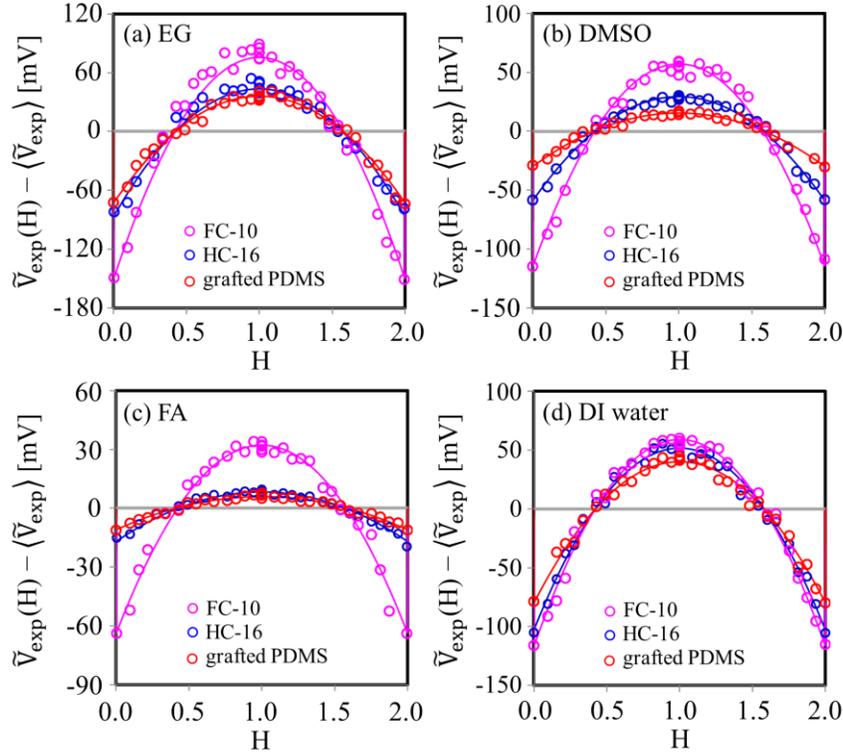

**Figure 4**. The mobility ($\tilde{v}_{exp}$) of the particles across the height (H) of a channel hydrophobized with fluorocarbon silane (pink), hydrocarbon silane (blue) and grafted PDMS (red). The experiment was carried out with (a) ethylene glycol (EG), (b) dimethyl sulfoxide (DMSO), (c) formamide (FA) and (d) DI water as test liquids. The curves are obtained from eq. 12 with the subtraction of $\langle \tilde{v}_{exp} \rangle$, as described in the text.

At the outset of the zeta potential analysis, we recognized that there are at least two unknown parameters: the slip length ($b$) and the Debye length ($\kappa^{-1}$). If these two parameters are known, the velocity profile can be fitted to equation 12 in order to extract the value of the zeta potential in a straightforward way. Before such a fit could be attempted, however, we also needed to address the possible uncertainty in the position of the particles (h') very close to the wall, which we expect to be somewhat larger than the radius of a single fluorescent particle. In our first attempt, we estimated the Debye lengths from the concentrations of the hydronium and hydroxide ions of the water present in these liquids (assuming a neutral pH) and those arising from the autoionization of the test liquid itself. The Debye lengths ($\kappa^{-1}$) were thus calculated to be 2 μm for ethylene glycol (autoionization constant = $10^{-16}$), 6 μm for formamide (autoionization constant = $10^{-17}$), 23 μm for DMSO, and 1 μm for pure water, respectively.[17] With these estimated Debye lengths, and with the help of a rather large electro-osmotic data set for each liquid, a multi-variable optimization was performed by adjusting the values of $b$, h', $D$ and $\zeta_w$ to find the best fit between the experimental data and equation 12. The correlation coefficients ($R^2$) of the best fits, however, had values less than 0.9 for all cases (e.g. 0.88 for water, 0.82 for EG, 0.59 for FA and 0.76 for DMSO on the hydrocarbon surface, and 0.90 for water, 0.85 for EG, 0.65 for FA and 0.84 for DMSO on the fluorocarbon surface, respectively). The actual concentrations of the hydronium ions in the non-aqueous liquids are, however, much smaller than the above estimates based on their pH (all greater than 9) measured with a standard pH meter. While these values are somewhat unreliable (the actual pH is even higher) owing to the fact that protons diffusing out of the electrode influence the measurement, they still produce rather large Debye length for all the liquids (10 μm for EG, 20 μm for FA and 500 μm for DMSO). These suggest that the autoionization of the residual water or the carrier liquids do not yield the correct values of the Debye length in any of the cases, including pure water. Next we performed a multi-variable optimized fitting of the experimental data by allowing the Debye length to be a variable as well. Such an analysis revealed that the slip length has to be vanishingly small, the Debye length has to be << 1 μm, and h' has to be less than 5 times the radius of the tracer particle (i.e., the particle is indeed very close to the walls of the channel) in order to obtain $R^2 > 0.98$. However, the lack of the precise knowledge of these parameters still makes it difficult to estimate the absolute values of the zeta potential. For a sufficiently thick channel, i.e., h >> b, and for a very short Debye length, the velocity profile of the tracer particle above an electrical screening layer can be written in the following form:

$$\tilde{V}_{exp}(H) = -\frac{\eta V_{exp}}{\varepsilon \varepsilon_0 E} = -\frac{D}{2}\left[h^2\left(H^2 - 2H\right)\right] + \zeta'_w + \left\langle \tilde{V}_{exp} \right\rangle \tag{14}$$

where, $\zeta'_w = \zeta_w - (b\kappa k_B T / qe)\sqrt{2\left[\cosh\left(A\zeta_w\right) - 1\right]}$ is the apparent zeta potential of the surface, the value of which would be same as the true zeta potential $\zeta_w$ if the fluid does not slip at the wall. Since the roughness ($\geq$0.6 nm) of both the hydrocarbon and fluorocarbon substrates are larger than the threshold roughness[28] of wall slippage, it is possible that the slip lengths (*b*) are vanishingly small in our experiments so that $\zeta''_w \sim \zeta_w$. We re-emphasize that the integrated values of $\tilde{V}_{exp}(H)$ for each liquid yielding the values of the mobility of the particle ($\mu_e$) are essentially the same on both the hydrocarbon and fluorocarbon surfaces, thus signifying that slippage of the liquids, if any, is not different on the two types of surface. Even when the values of $\mu_e$ differed slightly in different measurements, the estimated $\zeta''_w$ values were found to be rather robust. Within the above scenario, the apparent zeta potential $\zeta''_w$ of each liquid contacting either surface can be obtained by a straightforward fit of the experimental velocity data to equation 14 or can be taken from the intercept of the ordinates of Fig. 4 for each surface.

**Table 3**. Zeta potential ($\zeta'_w$) of glass surfaces modified with hydrocarbon silane (HC-16), fluorocarbon silane (FC-10) and grafted PDMS measured with different liquids.

| Liquid | Water Content (wt %) | Dielectric Constant[33-36] | Dipole Moment[36] (Debye) | Conductivity (µS/cm) | $\mu_e \times 10^{-9}$ (m²/V s) | | | $\zeta'_w$ (mV) | | |
|---|---|---|---|---|---|---|---|---|---|---|
| | | | | | HC-16 | FC-10 | grafted PDMS | HC-16 | FC-10 | grafted PDMS |
| EG | 0.03 | 38 | 2.31 | 0.23 ± 0.02 | 1.0 | 1.2 | 1.2 | -92 ± 3 | -144 ± 3 | -73 ± 2 |
| DMSO | 0.1 | 47 | 4.06 | 0.29 ± 0.01 | 5.0 | 4.1 | 7.0 | -54 ± 1 | -92 ± 2 | -30 ± 1 |
| FA | 0.07 | 110 | 3.37 | 151.3 ± 0.3 | 11.0 | 11.9 | 6.4 | -17 ± 1 | -66 ± 2 | -11 ± 1 |
| Water | 100 | 79 | 1.82 | 0.24 ± 0.02 | 45.9 | 53.4 | 50.9 | -102 ± 3 | -131 ± 3 | -80 ± 2 |
| Water | 100 | 79 | 1.82 | 0.18 ± 0.02 | 60.7 | 44.4 | --- | -103 ± 2 | -117 ± 3 | ---- |

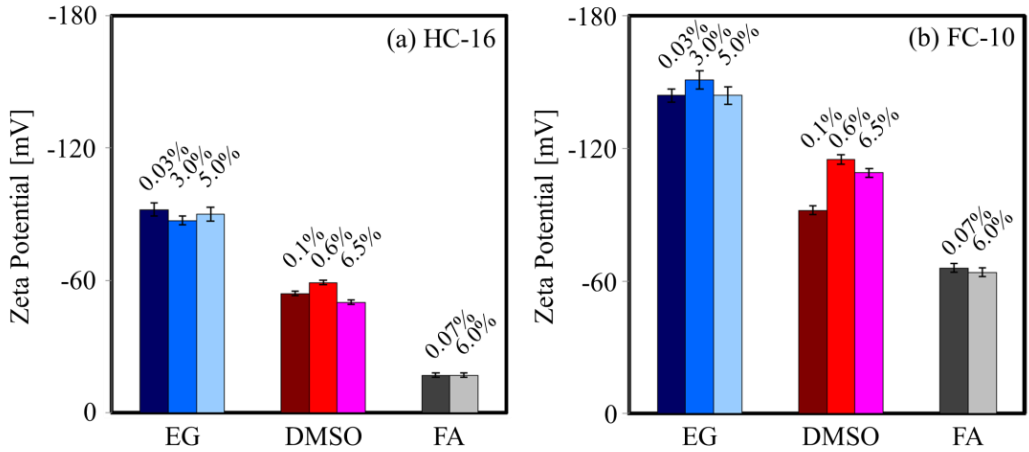

**Figure 5.** Bar graphs showing the zeta potentials of (a) HC-16 and (b) FC-10 silane modified glass tested against the ethylene glycol (EG), dimethyl sulfoxide (DMSO) and formamide (FA). The weight percent of water of each probe liquid is shown above the bars.

Table 3 summarizes the apparent zeta potentials estimated for each liquid against the hydrocarbon and the fluorocarbon surfaces. Significant zeta potential is indeed observed with each of the test liquids in contact with both of the hydrophobic surfaces, with the striking fact that its value on the fluorocarbon surface is substantially larger than that on the hydrocarbon one. From the very outset, we note that there is no correlation between the zeta potentials and the dipole moments of these liquids, which contrasts a model[32] suggesting that orientation of water dipoles at the interface contributes to the zeta potential.

We discussed above that the autoionization of the residual water or the carrier liquids do not yield the correct values of the Debye length needed to explain the electro-kinetic effects observed with any of the polar liquids. Additional measurements carried out with the HC-16 and FC-10 surfaces show that the zeta potentials of ethylene glycol, formamide or DMSO are, in fact, rather insensitive to the amount of residual water (Figure 5). For example, the zeta potential observed with ethylene glycol against a HC-16 surface remains around -90 mV even when the concentration of water in the solvent increases from 0.03 % to 5 % (w/w). Formamide and DMSO also show similar trends. The conductivity (151 µS/cm) of the as-received formamide used to test the zeta potential was rather high. However, when the liquid was treated with a mixed-bed resin, its conductivity decreased to 25 µS/cm, while its water content increased from 0.07% to 6% and pH decreased from 9.5 to 5.1. Remarkably, both the treated and the untreated formamide displayed very similar zeta potentials against HC-16 surface as shown in Figure 5. We expected the hydroxide ion concentration to increase with the dielectric constant of each probe liquid, and using the law of mass action, zeta potential to increase with the concentration of water as well. We thus find, so far, no clear evidence to support a model in which the ionization of the residual water or that of the carrier liquid plays a major role in the observed electro-kinetic phenomena.

As is usually the case with the surface chemistry experiments, it is tempting to attribute such types of anomalous results to impurities on the test surfaces that would deprotonate and give rise to surface charging. Since XPS did not provide evidence for the presence of such dissociable functional groups as carboxylic acid, one possibility is that the silanol groups of the support or from the silane used to modify the surfaces could be the source of such ionization. While such a picture would also be consistent with our findings (not reported here) and some observations reported in the literature[37,38] that the zeta potentials of the silanized silica surfaces increase with pH, it does not resolve some of the other issues satisfactorily. For example, let us take the case of formamide, which was used both as-received and after its treatment with a mixed-bed resin. Owing to the fact that the conductivity and the dielectric constant of the as-received formamide are both higher than those of water, this solvent provides a better environment for the putative silanol groups (isoelectric point ~ 2) to deprotonate than water. Thus one might expect formamide to display a higher surface charge density than water and thus a higher zeta potential, which is not the case. Furthermore, the zeta potentials of ethylene glycol, whose dielectric constant is considerably smaller than those of water and formamide, are quite comparable to that of water on the HC surface. For ethylene glycol on the FC surface, the zeta potential is actually slightly higher than water and nearly three times as large as that for formamide. Thus, the dissociation of a pre-existing group such as silanol to yield the negative charge does not appear to account entirely the trends of the zeta potentials observed with all the liquids against the various surfaces.

## 4.3. COMPARISON WITH THE GRAFTED PDMS FILM

Perhaps, a somewhat clearer picture emerges when we examine what happens when these polar liquids contact PDMS-coated glass slides for which the $Si_{2p}$ photoelectrons arising from glass are fully attenuated. The signals of the $Si_{2p}$ photoelectrons ejected from the 5-nm-thick grafted film correspond to the more electron-rich silicon atoms of PDMS, relative to those of silica. Furthermore, there is no evidence of any oxidized carbon species on such surfaces. The contact angles of the probe liquids also exhibit much lower hysteresis on this surface, thus indicating that these surfaces are rather homogeneous and devoid of pining sites of the types that give rise to contact angle hysteresis (Table 2).

Significant zeta potentials are observed with this surface as well (Fig. 4 and Table 3). Even though the magnitudes of the potentials on the PDMS-grafted surface are considerably smaller than the fluorocarbon coated glass slide, the values are only slightly smaller than the HC-16-coated glass slide, thus suggesting the effects observed with the HC-16 surface may indeed be mainly due to the hydrocarbon groups with a small effect arising from the underlying silica. The fluorocarbon monolayer, which can exhibit considerable disorder with a high area fraction of grain boundaries, can expose both silica as well as the acidic α-methylene group to H-bonding and donor-acceptor interactions with the probe liquids. What appears clear from these studies is that it is not imperative that a dissociative functional group pre-exist on a surface in order to give rise to charging in contact with polar liquids.

## 5. SUMMARY AND CONCLUSIONS

This research adds to the repertoire of observations[5,18,37-39] of electro-kinesis of water in contact with numerous hydrophobic surfaces. There is a common consensus in the literature that certain amount of charging does occur with any hydrophobic surface in contact with water. Finite zeta potentials have also been reported for hydrophobic self-assembled monolayers on gold[40-42], though there are large variations in the reported data depending upon the conditions used. For example, whereas Knoll *et al.*[40,41] reported a zeta potential in the range of -100 mV for aqueous electrolyte solutions in contact with flat surfaces of alkanethiol-modified gold, Yang and Abbott[42] reported a zeta potential of only 2 mV for alkanethiol-modified colloidal particles in water. While our observations do not lead to a conclusive picture of the origin of electro-kinesis at the interfaces of various polar liquids in contact with hydrodrophobic surfaces, they do illustrate how ubiquitous the phenomenon is even with some of the passive surfaces studied here. The strengths of the zeta potentials are so significant in these systems that it obligates us to be cautious in the quantitative interpretation of the zeta potentials measured even in more obvious situations involving ionizable functional groups.

We do not claim that our silanized glass surfaces present pure and homogeneous hydrophobic groups to the probe liquids. The finite wetting hysteresis observed with such surfaces suggests that there are defects. While the absolute values of the advancing and receding contact angles of water on our HC-16 treated glass surface are quite comparable to some of the carefully prepared monolayers[20] on silicon, there are reports[43,44] in the literature that monolayers of even lower hysteresis can be prepared depending upon the preparation condition, reactivity and the smoothness of the substrate.

However, it might be virtually impossible to produce a sufficiently defect-free monolayer coated surface of large enough surface area to conduct these measurements. Furthermore, the structures of such alkylsiloxane-coated substrates are known[45] to have a thin intercalated water layer. In spite of these complexities, the zeta potentials of various probe liquids in contact with the HC-16 treated glass surface are found to be rather close to those of a PDMS-grafted (5 nm) glass surface that is much more passive by wettability and XPS. The higher hysteresis of the fluorocarbon monolayer coated glass, however, suggests that it may not be all that passive in comparison to the other surfaces. Here, disorder and non-ideal surface coverage could indeed expose the underneath silica and the α-methylene group to the probe liquid. High zeta potentials of various probe liquids, including, aprotic DMSO, observed on surfaces suggest that electro-osmosis can be a very sensitive tool to study surface heterogeneity and defects.

Amongst all the surfaces studied here, the PDMS coating turned out to be most passive in terms of wettability, XPS and zeta potential. By extrapolating the results obtained with the HC and FC surfaces in terms of their insensitivity to the amount of the water present in the probe liquids, and that there is no evidence of any pre-existing functionality that would de-protonate, we feel that a non-conventional explanation would be needed to explain the charging of a passive surface as PDMS in contact with polar liquids.

We feel that the recent suggestions[6,14] of charging of hydrophobic surfaces in contact with water based on the transfer of charge inherent in hydrogen bonding deserve careful consideration. It has been well-known since the days of Mulliken[46] that hydrogen bonds between water molecules involve a transfer of electronic charge. According to this model[14], although the charge transfer is symmetrical in the bulk, the balance between the donating and accepting H-bonds is broken near a hydrophobic surface that leads to a net negative charge. Although certain details still need to be worked out[14] (e.g. the nature of the charged species that drag the liquid near the wall need to be established), this explanation of the charging at water/hydrophobic interfaces is promising and implies that other types of H-bonding liquids like ethylene glycol and formamide could show similar effects. However, in order to explain the differences with various surfaces, and especially the result obtained with the aprotic liquid (DMSO), additional factors need to be considered. The results with the aprotic DMSO on the FC surface is particularly striking, as this liquid exhibits a zeta potential as high as water or ethylene glycol, whereas its zeta potential against the HC and the PDMS surfaces are substantially lower than those two liquids. This result could be a manifestation of substantial donor-acceptor interaction between basic DMSO with the underneath silica and/or the acidic α-methylene of the fluorocarbon. In addition, a carbon-fluorine bond is considerably more polar than a carbon-hydrogen bond[47], which may further disrupt the charge transfer between the molecules of the hydrogen-bonding liquids. There is also the possibility that various probe liquids may themselves participate in donor-acceptor interactions with the fluorocarbon surface. Indeed, a fluorocarbon surface (e.g. Teflon) is more negative than a hydrocarbon surface (e.g. polyethylene) in the triboelectric series[16,48] as well. Recently, some interesting developments have taken place in the field of contact charging[49-52], where both ion and electron transfer have been considered. These views may be refined and blended with the picture proposed by Vácha et al[6] to improve our understanding of the zeta potentials observed with various protic and aprotic liquids against hydrophobic surfaces. The model also needs to be developed further in order to understand the results of Yaminsky and Johnston[15], who observed charging of a hydrophobized glass slide when it is retracted from water, formamide, and other liquids. When such a substrate is retracted from a liquid, it might emerge as a neutral species if the charge equilibrates quickly. However, in the presence of an energy barrier, such a substrate may emerge with a net charge, which will eventually equilibrate with atmosphere. Thus, it may also be necessary to invoke an activated intermediate state in the charge-transfer interaction.

Separation of such types of contact suggests the possibility that the contact and separation of a liquid and a solid involve irrecoverable work, which could be another hitherto unsuspected cause of contact-angle hysteresis in some situations. Further systematic work is warranted to resolve these issues. The observations of this work suggest the need to extend experimental and theoretical studies of electrokinetic phenomena beyond water and oil, to include a much larger spectrum of liquid-substrate combinations. The series of novel surfaces reported in reference [47], in which the headgroup properties of self-assembled monolayers can be varied systematically, may be valuable for such types of studies.

**ACKNOWLEDGEMENTS**

This work was supported by the Office of Naval Research (contract # N000140810743).

**REFERENCES**

1. Dickinson, W. The Effect of pH upon the Electrophoretic Mobility of Emulsions of Certain Hydrocarbons and Aliphatic Halides. *Trans. Faraday Soc.* **1941**, *37*, 140-148.

2. Marinova, K. G.; Alargova, R. G.; Denkov, N. D.; Velev, O. D.; Petsev, D. N.; Ivanov, I. B.; Borwankar, R. P. Charging of Oil-Water Interfaces Due to Spontaneous Adsorption of Hydroxyl Ions. *Langmuir* **1996**, *12*, 2045-2051.

3. Beattie, J. K.; Djerdjev, A. M. The Pristine Oil/Water Interface: Surfactant-Free Hydroxide-Charged Emulsions. *Angew. Chem. Int. Ed.* **2004**, *43*, 3568-3571.

4. Beattie, J. K.; Djerdjev, A. M.; Warr, G. G. The Surface of Neat Water is Basic. *Faraday Discuss.* **2009**, *141*, 31-39.

5. Tandon, V.; Bhagavatula, S. K.; Nelson, W. C.; Kirby, B. J. Zeta Potential and Electroosmotic Mobility in Microfluidic Devices Fabricated from Hydrophobic Polymers: 1. The Origins of Charge. *Electrophoresis* **2008**, *29*, 1092-1101.

6. Vácha, R.; Rick, S. W.; Jungwirth, P.; de Beer, A. G. F.; de Aguiar, H. B.; Samson, J.-S.; Roke, S. The Orientation and Charge of Water at the Hydrophobic Oil Droplet-Water Interface. *J. Am. Chem. Soc.* **2011**, *133*, 10204-10210.

7. Winter, B.; Faubel, M.; Vácha, R.; Jungwirth, P. Behavior of Hydroxide at the Water/Vapor Interface. *Chem. Phys. Lett.* **2009**, *474*, 241-247.

8. Petersen, P. B.; Saykally, R. J. Is the Liquid Water Surface Basic or Acidic? Macroscopic vs. Molecular-Scale Investigations. *Chem. Phys. Lett.* **2008**, *458*, 255-261.

9. Vácha, R.; Buch, V.; Milet, A.; Devlin, J. P.; Jungwirth, P. Autoionization at the Surface of Neat Water: Is the Top Layer pH Neutral, Basic, or Acidic? *Phys. Chem. Chem. Phys.* **2007**, *9*, 4736-4747.

10. Buch, V.; Milet, A.; Vácha, R.; Jungwirth, P.; Devlin, J. P. Water Surface Is Acidic. *Proc. Natl. Acad. Sci. U.S.A.* **2007**, *104*, 7342-7347.

11. Iuchi, S.; Chen, H. N.; Paesani, F.; Voth, G. A. Hydrated Excess Proton at Water-Hydrophobic Interfaces. *J. Phys. Chem. B* **2009**, *113*, 4017-4030.

12. Takahashi, H.; Maruyama, K.; Karino, Y.; Morita, A.; Nakano, M.; Jungwirth, P.; Matubayasi, N. Energetic Origin of Proton Affinity to the Air/Water Interface. *J. Phys. Chem. B* **2011**, *115*, 4745-4751.

13. Mundy, C. J.; Kuo, I. F. W.; Tuckerman, M. E.; Lee, H.-S.; Tobias, D. J. Hydroxide Anion at the Air-Water Interface. *Chem. Phys. Lett.* **2009**, *481*, 2-8.

14. Ben-Amotz, D. Unveiling Electron Promiscuity. *J. Phys. Chem. Lett.* **2011**, *2*, 1216-1222.

15. Yaminsky, V. V.; Johnston, M. B. Static Electrification by Nonwetting Liquids. Contact Charging and Contact Angles. *Langmuir* **1995**, *11*, 4153-4158.

16. Harper, W. R. *Contact and Frictional Electrification*; Laplacian Press: Morgan Hill, CA, **1998**.

17. Izutsu, K. *Electrochemistry in Nonaqueous Solutions*; Wiley-VCH: Weinheim, Germany, **2002**.

18. Lin, C.-H.; Chaudhury, M. K. Using Electrocapillarity to Measure the Zeta Potential of a Planar Hydrophobic Surface in Contact with Water and Nonionic Surfactant Solutions. *Langmuir* **2008**, *24*, 14276-14281.

19. Chaudhury, M. K.; Whitesides, G. M. Direct Measurement of Interfacial Interactions between Semispherical Lenses and Flat Sheets of Poly(dimethylsiloxane) and Their Chemical Derivatives. *Langmuir* **1991**, *7*, 1013-1025.

20. Wasserman, S. R.; Tao, Y.-T.; Whitesides, G. M. Structure and Reactivity of Alkylsiloxane Monolayers Formed by Reaction of Alkyltrichlorosilanes on Silicon Substrates. *Langmuir* **1989**, *5*, 1074-1087.

21. Ulman A. *An Introduction to Ultrathin Organic Films: From Langmuir-Blodgett to Self-Assembly*; Academic Press, Inc.: San Diego, CA, **1991**.

22. Krumpfer, J. W.; McCarthy, T. J. Contact Angle Hysteresis: A Different View and a Trivial Recipe for Low Hysteresis Hydrophobic Surfaces. *Faraday Discuss.* **2010**, *146*, 103-111.

23. Gao, C.; Xu, B.; Gilchrist, J. F. Mixing and Segregation of Microspheres in Microchannel Flows of Mono- and Bidispersed Suspensions. *Phys. Rev. E* **2009**, *79*, 036311.

24. Masliyah, J. H.; Bhattacharjee, S. *Electrokinetic and Colloid Transport Phenomena*; John Wiley & Sons, Inc.: Hoboken, NJ, **2006**, p 221-361.

25. Rothstein, J. P. Slip on Superhydrophobic Surfaces. *Annu. Rev. Fluid Mech.* **2010**, *42*, 89-109.

26. Hunter, R. J. *Zeta Potential in Colloid Science: Principles and Applications*; Academic Press, Inc.: London, U.K., **1986**, p 363-369.

27. Das, S.; Chakraborty, S. Effect of Conductivity Variations within the Electric Double Layer on the Streaming Potential Estimation in Narrow Fluidic Confinements. *Langmuir* **2010**, *26*, 11589-11596.

28. Zhu, Y. X.; Granick, S. Limits of the Hydrodynamic No-Slip Boundary Condition. *Phys. Rev. Lett.* **2002**, *88*, 106102.

29. Electrokinetic mobility is defined as the ratio of velocity to electric field. Since the mobility here is multiplied by the viscosity and divided by the dielectric constant of the liquid, $\tilde{V}_{exp}(H)$ is strictly speaking a depth dependent electrokinetic potential.

30. Mettu, S.; Chaudhury, M. K. Stochastic Relaxation of the Contact Line of a Water Drop on a Solid Substrate Subjected to White Noise Vibration: Roles of Hysteresis. *Langmuir* **2010**, *26*, 8131-8140.

31. Timmons, C. O.; Zisman, W. A. The Effect of Liquid Structure on Contact Angle Hysteresis. *J. Colloid Interface Sci.* **1966**, *22*, 165-171.

32. Wilson, M. A.; Pohorille, A.; Pratt L. R. Comment on ''Study on the Liquid-Vapor Interface of Water. I. Simulation Results of Thermodynamic Properties and Orientational Structure''. *J. Chem. Phys.* **1989**, *90*, 5211-5213.

33. *Ethylene Glycol Product Guide*; MEGlobal: Midland, MI.

34. Kaatze, U.; Pottel, R.; Schäfer M. Dielectric Spectrum of Dimethyl Sulfoxide/Water Mixtures as a Function of Composition. *J. Phys. Chem.* **1989**, *93*, 5623-5627.

35. Hernández-Luis, F.; Galleguillos-Castro, H.; Esteso, M. A. Activity Coefficients of NaF in Aqueous Mixtures with ε-Increasing Co-Solvent: Formamide-Water Mixtures at 298.15 K. *Fluid Phase Equilib.* **2005**, *227*, 245-253.

36. Riddick, J. A.; Bunger, W. B.; Sakano, T. K. *Organic Solvents: Physical Properties and Methods of Purification*, 4th ed.; John Wiley & Sons, Inc.: New York, **1986**.

37. Hozumi, A.; Sugimura, H.; Yokogawa, Y.; Kameyama, T.; Takai, O. ζ-Potentials of Planar Silicon Plates Covered with Alkyl- and Fluoroalkylsilane Self-Assembled Monolayers. *Colloids Surf. A* **2001**, *182*, 257-261.

38. Shyue, J.-J.; De Guire, M. R.; Nakanishi, T.; Masuda, Y.; Koumoto, K.; Sukenik C. N. Acid-Base Properties and Zeta Potentials of Self-Assembled Monolayers Obtained via in Situ Transformations. *Langmuir* **2004**, *20*, 8693-8698.

39. Kirby, B. J.; Hasselbrink, E. F. Zeta Potential of Microfluidic Substrates: 2. Data for Polymers. *Electrophoresis* **2004**, *25*, 203-213.

40. Schweiss, R.; Welzel, P. B.; Werner, C.; Knoll, W. Dissociation of Surface Functional Groups and Preferential Adsorption of Ions on Self-Assembled Monolayers Assessed by Streaming Potential and Streaming Current Measurements. *Langmuir* **2001,** *17,* 4304-4311.

41. Schweiss, R.; Welzel, P. B.; Werner, C.; Knoll, W. Interfacial Charge of Organic Thin Films Characterized by Streaming Potential and Streaming Current Measurements. *Colloids Surf. A* **2001**, *195*, 97-102.

42. Yang, Z. Q.; Abbott, N. L. Spontaneous Formation of Water Droplets at Oil-Solid Interfaces. *Langmuir* **2010**, *26*, 13797-13804.

43. Maoz, R.; Sagiv, J. On the Formation and Structure of Self-Assembling Monolayers. I. A Comparative ATR-Wettability Study of Langmuir-Blodgett and Adsorbed Films on Flat Substrates and Glass Microbeads. *J. Colloid Interface Sci.* **1984,** *100*, 465-496.

44. Kessel, C. R.; Granick, S. Formation and Characterization of a Highly Ordered and Well-Anchored Alkylsilane Monolayer on Mica by Self-Assembly. *Langmuir*, **1991**, *7*, 532-538.

45. Angst, D. L.; Simmons, G. W. Moisture Absorption Characteristics of Organosiloxane Self-Assembled Monolayers, *Langmuir*, **1991**, *7*, 2236-2242.

46. Mulliken, R. S.; Person, W. B. Donor-Acceptor Complexes. *Annu. Rev. Phys. Chem.* **1962**, *13*, 107-126.

47. Graupe, M.; Takenaga, M.; Koini, T.; Colorado, R.; Lee, T. R. Oriented Surface Dipoles Strongly Influence Interfacial Wettabilities. *J. Am. Chem. Soc.* **1999**, *121*, 3222-3223.

48. Williams, M. W. Triboelectric Charging of Insulating Polymers–Some New Perspectives. *AIP Adv.* **2012**, *2*, 010701.

49. Gibson, H. W.; Bailey, F. C. Linear Free Energy Relationships. Triboelectric Charging of Poly(olefins). *Chem. Phys. Lett.* **1977**, *51*, 352-355.

50. Baytekin, H. T.; Baytekin, B.; Soh, S.; Grzybowski, B. A. Is Water Necessary for Contact Electrification? *Angew. Chem. Int. Ed.* **2011**, *50*, 6766-6770.

51. Wiles, J. A.; Grzybowski, B. A.; Winkleman, A.; Whitesides, G. M. A Tool for Studying Contact Electrification in Systems Comprising Metals and Insulating Polymers. *Anal. Chem.* **2003**, *75*, 4859-4867.

52. McCarty, L. S.; Whitesides, G. M. Electrostatic Charging Due to Separation of Ions at Interfaces: Contact Electrification of Ionic Electrets. *Angew. Chem. Int. Ed.* **2008**, *47*, 2188-2207.

For "Table of Contents Use Only"

# Electro-kinetics of Polar Liquids in Contact with Non-Polar Surfaces

Chih-Hsiu Lin[1], Gregory S. Ferguson[2] and Manoj K. Chaudhury*[1]

[1] Department of Chemical Engineering and [2] Department of Chemistry, Lehigh University, Bethlehem, PA 18015

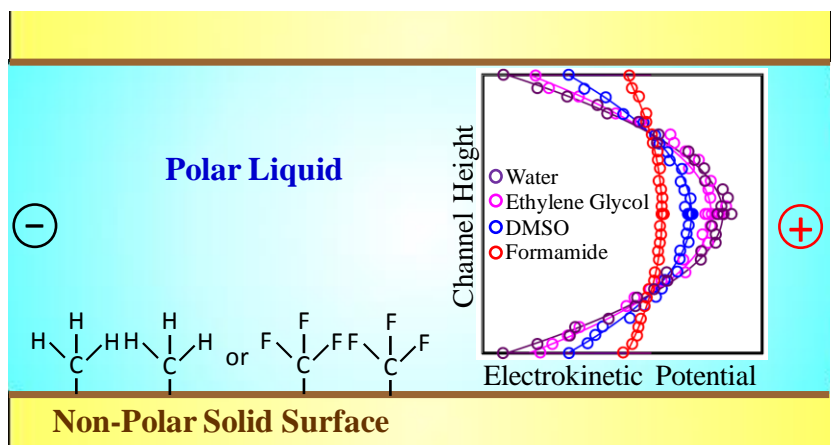